# Nonprobability follow-up sample analysis: an application to SARS-CoV-2 infection prevalence estimation

Yan Li[1], Laura Yee[2], Sally Hunsberger[3], Matthew J. Memoli[4], Kaitlyn Sadtler[5], Barry Graubard[6]

[1.] Joint Program in Survey Methodology & Department of Epidemiology and Biostatistics, University of Maryland at College Park. [2.] Division of Cancer Treatment and Diagnosis, National Cancer Institute (NCI), NIH. [3.] Division of Clinical Research, National Allergy and Infectious Disease Institute (NIAID), NIH. [4] LID Clinical Studies Unit, Laboratory of Infectious Diseases, NIAID, NIH. [5.] Section on Immuno-Engineering, National Institute of Biomedical Imaging and Bioengineering, NIH. [6.] Division of Cancer Epidemiology and Genetics, NCI, NIH.

## Abstract

Public health policy makers are faced with making crucial decisions rapidly during infectious disease outbreaks such as that caused by SARS-CoV-2. Ideally, rapidly deployed representative health surveys could provide needed data for such decisions. Under the constraints of a limited timeframe and resources, it may be infeasible to implement random based (probability) sampling that yields a population representative survey sample with high response rates. As an alternative, a volunteer (nonprobability) sample is often collected using outreach methods such as social media and web surveys. Compared to a probability sample, a nonprobability sample is subject to selection bias. In addition, when participants are followed longitudinally nonresponse often occurs at later follow up timepoints. As a result, estimates of cross-sectional parameters at later timepoints will be subject to selection bias and nonresponse bias. In this paper, we create kernel-weighted pseudoweights (KW) for the baseline survey participants and construct nonresponse-adjusted kw (kwNR) for respondents at each follow-visit to estimate the population mean at the follow-up visits. We develop Taylor Linearization variance estimation that accounts for variability due to estimating both pseudoweights and the nonresponse adjustments. Simulations are conducted to evaluate the proposed kwNR-weighted estimates. We investigate covariate effects on each of the following: baseline sample participation propensity, follow-up response propensity and the mean of the outcome. We apply the proposed kwNR-weighted methods to the SARS-CoV-2 antibody seropositivity longitudinal study, which begins with a baseline survey early in the pandemic, and collects data at six- and twelve-month post baseline follow-ups.

## 1. INTRODUCTION

The National Institutes of Health (NIH) seroprevalence study (ClinicalTrials.gov NCT04334954) began with a baseline survey at an early stage of the global COVID-19 pandemic (April 2020). The objective was to estimate the proportion of the US adults with anti-SARS-CoV-2 antibodies that had no confirmed history of SARS-CoV-2 infection. The pandemic environment led to immense nationwide interest in recruitment for the study as reflected by more than 460,000 volunteers responding within weeks of the study announcement. The number of volunteers was too large to use in its entirety because each study participant was required to provide a blood sample that needed laboratory seropositivity assessment. Therefore, a subsample was selected to reflect the US population based on the following six demographic variables: age group, sex, geographic region, race, ethnicity, and population density. As the subsample was selected from volunteers, we refer to it as a quota sample where the demographic variables used to select the subsample are the quota variables. The sample of individuals selected from the volunteer population were given a questionnaire and asked to provide a blood sample through home sampling kits. The invitation acceptance rate (# accepting the invitation / # of the total invitations = 11,382/27,716) was approximately 41%, and 29% (=8,058/27,716) of the total participants invited completed the survey items and provided baseline blood samples. From this data set we estimated that by July of 2020 at least 4.6% of the US population had previously been infected with COVID-19 without their prior knowledge, which was fivefold higher than the diagnosed rate at this early stage of the pandemic (Kalish et al. 2021).

The lives of people around the world have been dramatically impacted by the SARS-CoV-2 pandemic (Maxmen, 2021). Monitoring the prevalence of people who had antibodies to SARS-CoV-2 remained a question of interest throughout the pandemic, especially during the rollout and distribution of vaccines. The SARS-CoV-2 seroprevalence study presented the opportunity to further examine this question by following the original group of participants longitudinally. Respondents from the baseline survey (n= 8,058) received a follow-up web survey and home blood microsampling kit at six and twelve months after baseline. Participants who returned the blood samples were defined as respondents. The study retained 4,562 respondents at month 6 (visit 2) and 4,226 respondents at month 12 (visit 3). The respective response rates were 56.6% and 52.4% for month-6 and month-12 follow-up surveys (Figure 1).

Although the quota sample produced a random sample with known probabilities of selection from the pool of volunteers (Kalish, et al., 2021, Supplement, Section 4) for the baseline survey, the pool of more than 460,000 volunteers was a nonrandom sample of the targeted US population. This nonrandom sample may introduce *selection bias* that was not accounted for by the quota sampling. Furthermore, data collected in the follow-up surveys can result in attrition (Bolger and Laurenceau, 2013) and response rates are often found to be 10% or lower (Baker et al. 2013). Although low response rates do not necessarily indicate the presence of nonresponse bias (Groves and Peytcheva 2008; Brick and Tourangeau 2017), the potential for such bias is a significant concern as the composition of the follow-up samples often differs from that of the underlying population. For instance, in the seroprevalence study's follow-up surveys, a distinct discrepancy was observed between nonresponse rates in rural and urban areas. Therefore, when analyzing data from longitudinal surveys it is crucial to account for potential selection and nonresponse biases.

Research has been widely conducted to address these two potential biases. To reduce the *selection* bias in non-random samples (also called non-probability samples) such as quota samples, propensity score based (PS-based) weighting methods (Wang, et al., 2021; Chen, et al., 2020; Elliott & Valliant, 2017) or PS-based matching methods (Rivers, 2007; Wang, et al., 2020; Kern, et al., 2021) have been developed that provide approaches to approximate population-based inference. PS-based methods construct pseudoweights from the PS's for nonprobability samples to weight the distribution of variables in the nonprobability sample to be approximately the same as those in the sample-weighted reference (representative) survey, assuming the propensity model is correctly specified. For example, to reduce the *selection* bias in the COVID baseline sample, kernel weighted pseudoweights (KW), which are a type of pf PS-based matching method (Wang et al. 2020), are constructed using a representative survey (BRFSS 2018) as the reference (Kalish, et al. 2021). The pseudoweighted prevalence of seropositivity and its variance are estimated to account for the sampling and PS-based matching (Wang et al. 2021). To reduce the *nonresponse bias*, methods have been studied to estimate the propensity to respond to a survey (Groves 2006; Iannacchione et al. 1991) and to use these estimated response propensities to adjust sampling weights in representative surveys. The best (auxiliary) variables to be included for nonresponse propensity estimation are those that are correlated with the key outcomes of interest and secondarily correlated with the nonresponse process (Little and Vartivarian 2005). Although the literature has extensive statistical methods for reducing either the selection bias or the nonresponse bias, no statistical methods have been developed for simultaneously reducing both the selection bias and nonresponse bias in the analysis of nonprobability longitudinal survey samples.

The purpose of this paper is to present a statistical method to make population-based unbiased inferences using nonprobability longitudinal survey samples subject to both the potential selection bias and the nonresponse bias. We first construct the PS-based KW for the baseline nonprobability sample, then create nonresponse-adjusted KW pseudoweights (kwNR) for the follow-up nonprobability samples. The Taylor Linearization (TL) variance estimator of kwNR-weighted sample means is developed, accounting for the variability due to estimating both the pseudoweights and the response propensities. The rest of this paper proceeds as follows. In section 2, the kwNR estimator and its TL variance are developed. Simulations results evaluating the performance of the estimators are presented in section 3. Finally, we discuss our findings.

## 2. METHODS

### 2.1 Notation

Consider a target finite population (FP) as a random sample of *N* individuals from a superpopulation model, indexed by $U = \{1,2,\ldots,N\}$, with observations on a study variable $y$ of interest (e.g., SARS-CoV-2 antibody seropositivity) and on a vector of covariates $\boldsymbol{x}_i$. Let $\{y_i, \boldsymbol{x}_i : i \in C\}$ be the observations for the volunteer nonprobability sample of individuals that is collected from the baseline survey, where $C \subset U$ with size $n_c$. We are interested in estimating the finite population mean

$$\bar{Y}_N = \frac{\sum_{i \in U} y_i}{N}.$$

Using the nonprobability sample *C*, the population mean can be estimated using various propensity score (PS) based pseudoweighting methods, given by

$$\hat{\bar{Y}}^{kw} = \frac{\sum_{j \in C} \hat{d}_j y_i}{\sum_{i \in C} \hat{d}_j},$$

where $\hat{d}_j$ is the pseudoweight for the $j^{th}$ nonprobability sample individual that is an estimate of an implicit sample weight $d_j$ that is unobserved. In this paper, we use the kernel-weighted PS matching pseudoweights (KW) for the nonprobability sample *C* that was employed in Kalish et al. (2021), given by

$$\hat{d}_j = \sum_{i \in S} d_i K_{ij} \text{ with } K_{ij} = \frac{K\left(\frac{b(\boldsymbol{x}_i) - b(\boldsymbol{x}_j)}{h}\right)}{\sum_{l \in C} K\left(\frac{b(\boldsymbol{x}_i) - b(\boldsymbol{x}_l)}{h}\right)} \text{ for } j \in C,$$

where $d_i$ denotes the observed sampling weight for unit *i* in a reference probability sample, denoted by *S*, which is representative of the target FP. The $K(\cdot)$ is an arbitrary kernel function such as standard normal density function, $h$ is the bandwidth associated with $K(\cdot)$, and $b(\boldsymbol{x})$ is a balancing score, which satisfies the conditional exchangeability (CE) assumption, $E(y|b(\boldsymbol{x}), C) = E(y|b(\boldsymbol{x}), U)$, where the expectation on the right side is over the superpopulation model randomness of *y* in *U*, while the expectation of *y* in *C* on the left side is $E(y|b(\boldsymbol{x}), C) = E_U[E_C(y|b(\boldsymbol{x}), U)|b(\boldsymbol{x})]$, where the subscripts *C* and *U* refer to the expectation with respect to the unknown random nonprobability sample participation process from *U* and the superpopulation model randomness, respectively (Li 2023). The CE assumption is that given *b*(***x***), the outcome variable has the same expectation in *C* as in the FP. Under the CE assumption, it has been shown that the KW-weighted sample mean $\hat{\bar{Y}}^{kw}$ is design-consistent under certain regularity conditions (Wang, et al. 2021). Li (2022) discussed different methods for estimating balancing scores *b*(***x***). For example, Wang et al. (2021) assumes a logistic regression model,

$$\log\left\{\frac{p(\boldsymbol{x}_i)}{1 - p(\boldsymbol{x}_i)}\right\} = \alpha + \boldsymbol{B}^T g(\boldsymbol{x}_i) \quad \text{for } i \in U, \qquad (1)$$

where the propensity score $p(\boldsymbol{x}_i)$ is the propensity of individual *i* in FP being in the nonprobability sample versus the FP; $g(\boldsymbol{x}_i)$ is a known function of observed covariates; and $\alpha$ and $\boldsymbol{B}$ the unknown regression coefficients to be estimated. The balancing score was estimated by $b(\boldsymbol{x}; \widehat{\boldsymbol{B}}_w) = \widehat{\boldsymbol{B}}_w^{\boldsymbol{T}} g(\boldsymbol{x}_i)$ with $\widehat{\boldsymbol{B}}_w$ the estimates of ***B*** by fitting (1) to the combined nonprobability sample *C* and $d_i$-weighted representative probability sample *S*.

The difference in the balancing score $b(\boldsymbol{x}_i) - b(\boldsymbol{x}_j)$ measures the similarity in $\boldsymbol{x}$ distribution between the nonprobability sample unit $j \in C$ and the survey unit $i \in S$ (Rosenbaum and Rubin 1983). The KW pseudoweight $\hat{d}_j$ is the sum of probability sample weights $d_i; i \in S$, weighted by the kernel weight $K_{ij}$. Intuitively, the closer $b(\boldsymbol{x}_j)$ is to $b(\boldsymbol{x}_i)$, the higher similarity in the distribution of ***x*** and therefore, a larger portion of the probability sample weight $d_i$ is assigned to the nonprobability sample unit *j*.

**2.2 Kernel-weighted estimator with nonresponse adjustment (kwNR)**

We denote $R$, $R \subset C$, as the index set of respondents at a follow-up timepoint of sample size $n_r$.

$$\hat{\bar{Y}}^{kwNR} = \frac{\sum_{i \in C}(\hat{d}_i \hat{r}_i^{-1}) I_i^R y_i}{\sum_{i \in C}(\hat{d}_i \hat{r}_i^{-1}) I_i^R},$$

where $\hat{d}_i$ is the baseline KW pseudoweight, $I_i^R$ is responding indicator (=1 if $i \in R$; 0 otherwise), $\hat{r}_i = expit(\hat{\gamma}_0 + \hat{\boldsymbol{\gamma}}^T g(\boldsymbol{z}_i))$ is the estimate of $r_i = r(\boldsymbol{z}_i)$, the response propensity for unit *i*, assuming a logistic regression model

$$\log\left\{\frac{r(\boldsymbol{z}_i)}{1 - r(\boldsymbol{z}_i)}\right\} = \gamma_0 + \boldsymbol{\gamma}^T g(\boldsymbol{z}_i), \qquad \text{for } i \in C, \qquad (2)$$

where $\boldsymbol{z}_i$ is a set of observed covariates that are predictive of response propensity and/or the outcome (Little and Vartivarian 2005), which may include the same or different set of covariates as $\boldsymbol{x}_i$ in equation (1). The estimates $(\hat{\gamma}_0, \hat{\boldsymbol{\gamma}})$ is derived from solving for $(\gamma_0, \boldsymbol{\gamma})$ by maximizing the pseudo-loglikelihood function

$$\log L(\boldsymbol{\gamma}) = \sum_{i \in C} \hat{d}_i^* \{ I_i^R \log r_i + (1 - I_i^R) \log(1 - r_i) \}$$

with the estimating equation

$$S(\boldsymbol{\gamma}) = \sum\nolimits_{i \in C} \hat{d}_i^* (I_i^R - r_i) \boldsymbol{z}_i = 0$$

and pseudo-information matrix

$$\boldsymbol{I} = -\frac{\partial}{\partial \boldsymbol{\gamma}} S(\boldsymbol{\gamma}) = \sum\nolimits_{i \in C} \{ \hat{d}_i^* r_i (1 - r_i) \boldsymbol{z}_i \boldsymbol{z}_i^T \}.$$

The resulting $\boldsymbol{\gamma}$ estimates are denoted by $\hat{\boldsymbol{\gamma}}$. Note $\hat{d}_i^*$ can be baseline KW pseudoweights $\hat{d}_i^* = \hat{d}_i$ or common value of one $\hat{d}_i^* = 1$ without considering the baseline KW (Little and Vartivarian 2003).

### 2.3 Taylor linearization (TL) variance of the kwNR estimator

We derive the variance of $\hat{\bar{Y}}^{kwNR}$ that accounts for the variability due to estimating the participation propensity in $\hat{d}_i$ for the baseline sample as well as the response propensity $\hat{r}_i$ for the follow-up sample.

The NR-adjusted mean $\bar{y}_R$ by Taylor expansion is expressed by

$$\hat{\bar{Y}}^{kwNR} = \frac{\sum_{i \in C} (\hat{d}_i \hat{r}_i^{-1}) I_i^R y_i}{\sum_{i \in C} (\hat{d}_i r_i^{-1}) I_i^R} \approx \bar{Y}_N + \sum_{i \in C} \left( \frac{\partial}{\partial I_i^C} \hat{\bar{Y}}^{kwNR} \right) (I_i^C - \pi_i) + \sum_{i \in C} \left( \frac{\partial}{\partial I_i^R} \hat{\bar{Y}}^{kwNR} \right) (I_i^R - r_i),$$

where $\frac{\partial}{\partial I_i^C} \hat{\bar{Y}}^{kwNR}$ and $\frac{\partial}{\partial I_i^R} \hat{\bar{Y}}^{kwNR}$ are partial derivatives of $\hat{\bar{Y}}^{kwNR}$ with respect to the participant indicator $I_i^C$ and the respondent indicator $I_i^R$ for each unit $i \in C$. By doing this, we define the Taylor deviate (TD) by $\delta_l^R = \frac{\partial}{\partial I_l^R} \hat{\bar{Y}}^{kwNR}$ and $\delta_l^C = \frac{\partial}{\partial I_l^C} \hat{\bar{Y}}^{kwNR}$ for unit $l \in C$ where the TD are used to obtain a delta method expression for the variance estimator of $\hat{\bar{Y}}^{kwNR}$.

The variance of the NR-adjusted mean $\bar{y}_R$ is approximated by

$$Var\left(\hat{\bar{Y}}^{kwNR}\right) = Var\left[ E\left( \hat{\bar{Y}}^{kwNR} \middle| C \right) \right] + E\left[ Var\left( \hat{\bar{Y}}^{kwNR} \middle| C \right) \right],$$

where the first component of the right side can be estimated by

$$var_1 = var\left( \sum_{i \in U} I_l^C \delta_l^C \right),$$

with

$$\delta_l^C = \frac{\partial}{\partial I_l^C} \bar{y}_R = \frac{\hat{d}_l \hat{r}_l^{-1} I_l^R (y_l - \bar{y}_R) \{ 1 - (1 - \hat{r}_l) \boldsymbol{z}_l \sum_{i \in C} \hat{d}_i^* (I_i^R - \hat{r}_i) \boldsymbol{z}_i \hat{\boldsymbol{I}}^{-1} \}}{\sum_{i \in C} \hat{d}_i (\hat{r}_i^{-1} I_i^R)} \qquad (3)$$

and

$$\hat{\boldsymbol{I}} = \sum\nolimits_{i \in C} \hat{d}_i^* \{ \hat{r}_i (1 - \hat{r}_i) \boldsymbol{z}_i \boldsymbol{z}_i^T \}.$$

The $var(\sum_U I_l^C \delta_l^C)$ can be derived following the method of Wang et al. (2021); this accounts for (sampling) variability of $I_l^C$ in the estimation of the KW weights $\hat{d}_l$ and response propensity $\hat{r}_l$.

The second component is estimated by

$$var_2 = var(\sum\nolimits_{i \in C} I_l^R \delta_l^R \, | C),$$

where

$$\delta_l^R = \frac{\partial}{\partial I_l^R}\bar{y}_R = \frac{\hat{d}_l\hat{r}_l^{-1}(y_l - \bar{y}_R) - \left(\hat{d}_l^*\boldsymbol{z}_l\hat{\boldsymbol{I}}^{-1}\right)\sum_c \hat{d}_i\hat{r}_i^{-1}(1-\hat{r}_i)I_i^R\boldsymbol{z}_i(y_i - \bar{y}_R)}{\sum_{i\in C}\hat{d}_i\hat{r}_i^{-1}I_i^R}. \quad (4)$$

Assuming simple random sampling, $var_2$ can be estimated by,

$$var_2 \approx n_R\left(1 - \frac{n_R}{n_C}\right)var\left(\delta_l^R\right).$$

Finally, $\mathrm{Var}\left(\hat{\hat{Y}}^{kwNR}\right)$ is approximated by

$$var_{TL}\left(\hat{\hat{Y}}^{kwNR}\right) = var_1 + var_2.$$

### 2.4 Subgroup Estimation

For estimation of subgroups, e.g., seropositivity estimation for each of six regions, we define

$$\hat{\hat{Y}}_D^{kwNR} = \frac{\sum_{i\in C}\left(\hat{d}_i\hat{r}_i^{-1}\right)I_i^R\left(I_i^D y_i\right)}{\sum_{i\in C}\left(\hat{d}_i\hat{r}_i^{-1}\right)I_i^R I_i^D},$$

where $I_i^D = 1$ if unit *i* belongs to subgroup D; 0 otherwise. The variance of $\hat{\hat{Y}}_D^{kwNR}$ is derived along the same derivation for $Var(\hat{\hat{Y}}^{kwNR})$ in Section 2.3. The resulting subgroup variance estimates are

$$var\left(\hat{\hat{Y}}_D^{kwNR}\right) = var_1 + var_2,$$

where $\delta_l^C$ and $\delta_l^R$ are defined in (3) and (4), respectively, but with their denominators replaced by $\sum_{i\in C}\left(\hat{d}_i\hat{r}_i^{-1}\right)I_i^R I_i^D$.

## 3. SIMULATIONS

In the simulation, we first construct the KW pseudoweights for the nonprobability sample and the NR-adjusted KW (kwNR) for respondents in the follow-up surveys to estimate the population means using both the nonprobability baseline sample and the follow-up samples. The proposed Taylor Linearization variance estimator accounts for the variability due to estimation of the participation propensity and response propensity. We compare the KW-weighted with the kwNR-weighted respondent sample estimates in terms of relative bias and variance. Different covariate effects on 1) participation propensity, 2) response propensity, and 3) mean of the outcome are evaluated.

**Population Generation**

We generate a FP of size N=200,000 with one covariate $x$ following a standard normal distribution $N(0,1)$. The binary outcome Y is generated with the mean defined by

$$P(Y=1) = \frac{\exp\left(\beta_{y0} + x\beta_{y1}\right)}{1 + \exp\left(\beta_{y0} + x\beta_{y1}\right)}, \quad (3)$$

where $\beta_y = \left(\beta_{y0}, \beta_{y1}\right)^T$ are the model parameters specified as $\beta_{y0} = -0.5$ and $\beta_{y1} = 0.5$, so that the mean outcome is 38%, similar to the seropositivity rate in the real data example (Section 4) at visit 2 (six-month follow-up). Each FP individual has a propensity of responding to the survey. A binary responding status (R=1 if responding; 0 otherwise) is generated with the mean defined by $P(I^R = 1) = \frac{\exp(\beta_{r0} + x\beta_{r1})}{1+\exp(\beta_{r0} + x\beta_{r1})}$ with $\beta_{r0} = 0.2$ and $\beta_{r1} = 0.5$ that results in an average response rate of 55%. We also vary the value of $\beta_{y0} = 2$ and $\beta_{r0} = 1$ to have the mean outcome 88% and the response rate 73% to study their effects on the performance of the proposed kwNR estimator. The results show similar pattern, and therefore, are presented under $\beta_{y0} = 0.5$ and $\beta_{r0} = 0.2$ only.

We generate a complete dataset for the FP of size N=200,000, consisting of one covariate *x*, a binary outcome *Y*, and a binary responding status *R* at a follow-up time, denoted by {$(X_i, Y_i, I_i^R)$; for *i* = 1,…,N}.

.

**Nonprobability and Probability Sample Selection**

To reflect the COVID baseline sample, a nonprobability sample of size $n_c = 8{,}000$, denoted by $C$, is selected from the FP using probability proportional to size (PPS) sampling and the measure of size $(mos_k)$ for the $k^{th}$ FP individual is defined by

$$mos_k = \exp(\beta_{c0} + x_k\beta_{c1})$$

so that the inclusion probability for the $k^{th}$ FP individual (Korn and Graubard, 1999) is

$$P(k \in C|U) = \frac{n_c \times mos_k}{\sum_{k\in U} mos_k}.$$

The parameters for inclusion probabilities are varied, $\beta_c = (\beta_{c0}, \beta_{c1}) = (-1, 1.5)$ and $(-1, 0.5)$, so that the coefficient of variation (CV) of the sample weights $(d_k; k \in C)$, defined by the inverse of the inclusion probability, are $CV(d_k; k \in C) = 2.55$ and 0.52, respectively. In the COVID sample, all design variables such as inclusion probabilities are unobservable. The sample weights $(d_k; k \in C)$ need to be estimated in the real-life analysis and therefore we estimate sample weights for *C* units by the KW method in the simulations. A probability sample of size $n_s = 2000$, denoted by $S$, is selected by simple random sampling, which is FP representative.

We first construct KW pseudo-weights for the nonprobability sample $C$ by fitting the combined sample (i.e., the nonprobability and probability sample) to the logistic propensity model (Wang, et al. 2022). We denote the mean estimates using *C* units by KW methods and their estimated standard errors (SE's) as $\hat{\bar{Y}}_c^{kw}$ and $se_c^{kw}$, respectively. In the follow-up survey, the outcome is observable from respondents only. We conduct nonresponse (NR) adjustment by multiplying the KW pseudoweights by the inverse of estimated response propensity as the new NR-adjusted KW weights, denoted by kwNR. Additional variability due to the estimation of the NR adjustment are accounted for in the SE of the mean estimates using respondents only.

**Simulation results**

Table 1 gives the results of the simulations. We denote the proposed follow-up survey estimates adjusting for NR and their estimated Taylor linearization standard error by $\hat{\bar{Y}}_r^{kwNR}$ and $se_r^{kwNR}$, respectively. For comparison purposes, we also report the mean estimates using the nonprobability sample C with true sample survey weights ($\hat{\bar{Y}}_c^d$), without weights ($\hat{\bar{Y}}_c^0$), with KW weights but without NR adjustments ($\hat{\bar{Y}}_c^{kw}$), or using the respondent sample R without weights ($\hat{\bar{Y}}_r^0$), and with KW weights but without NR adjustments ($\hat{\bar{Y}}_r^{kw}$).

The relative bias (RB = bias divided by the population mean × $100\%$ with the bias defined by the difference between the mean of 2,000 simulated means and the population mean), empirical variance (empVar= variance of 2,000 simulated means), mean squared error (MSE = square of bias + empirical variance), and variance ratio (VR = average of 2,000 TL variance estimates /empVar) are reported.

Table 1. Population mean estimation with varying covariate effects on participation propensity when $p(Y = 1) = 38\%$ for 2000 simulations

| Estimator | Relative Bias × $10^2$ | | Emp Var × $10^4$ | | MSE × $10^4$ | |
|---|---|---|---|---|---|---|
| | $CV$(KW)=0.52 | $CV$(KW)=2.55 | $CV$(KW)=0.52 | $CV$(KW)=2.55 | $CV$(KW)=0.52 | $CV$(KW)=2.55 |
| | Nonprobability Baseline Sample Estimates | | | | | |
| true-weighted $\hat{\bar{Y}}_c^d$ | -0.34 | -1 | 0.34 | 2.24 | 0.36 | 2.39 |
| un-weighted $\hat{\bar{Y}}_c^0$ | 14.66 | 40.44 | 0.29 | 0.27 | 31.85 | 240.33 |
| kw-weighted $\hat{\bar{Y}}_c^{kw}$ | 0.26 | 0.77 | 0.39 | 1.85 | 0.4 | 1.94 |
| | Respondent Follow-up Sample Estimates | | | | | |
| un-weighted $\hat{\bar{Y}}_r^0$ | 20.27 | 44.11 | 0.49 | 0.38 | 60.85 | 286.02 |
| kw-weighted $\hat{\bar{Y}}_r^{kw}$ | 6.33 | 6.59 | 0.63 | 2.62 | 6.52 | 8.99 |
| kwNR-weighted $\hat{\bar{Y}}_r^{kwNR}$ | 0.18 | 0.72 | 0.75 | 4.84 | 0.76 | 4.91 |

Table 1 presents results with varying $\beta_{c1} = 0.5$ and 1.5, corresponding to the CV of baseline KW pseudoweights of (CV(KW)=0.52) and large (CV(KW)=2.55), respectively. The prevalence of the outcome is on average 38% with $\beta_{y0} = -0.5$. Three observations are made. *First*, ignoring the selection bias, the unweighted baseline sample (denoted by subscript *C*) estimates $\left(\hat{\bar{Y}}_c^0\right)$ are consistently biased with relative biases of 14.66% and 40.44%; whereas KW-weighted baseline sample estimates $\left(\hat{\bar{Y}}_c^{kw}\right)$ reduce the selection bias with relative biases close to zero (0.26% and 0.77%). Using the follow-up respondent sample (denoted by the subscript *R*) that is subject to both the selection bias and nonresponse bias, unweighted estimates ($\bar{y}_r^0$) are badly biased with relative biases of 20.27% and 44.11%. To remedy this, two sets of pseudoweights are applied. It can be observed that KW-weighted follow-up samples corrected for the selection bias with reduced relative bias of 6.33% and 6.59%, whereas the proposed kwNR-weighted follow-up sample estimates ($\hat{\bar{Y}}_r^{kwNR}$), which corrects for both the selection bias and the nonresponse bias, are approximately unbiased with relative biases close to zero (0.18% and 0.72%). *Secondly*, unweighted baseline or follow-up sample estimates are efficient. In contrast, the proposed kwNR-weighted estimates have the largest empirical variances. These results are as expected since the set of pseudoweights (kwNR) is constructed based on estimated participation and response propensities, which therefore is variable and produces more variable estimates than unweighted estimates. *Lastly*, combining the first two observations, the MSE of the kwNR-weighted follow-up sample means are the smallest among the three follow-up sample estimates, i.e., MSE($\hat{\bar{Y}}_r^{kwNR}$) < MSE($\hat{\bar{Y}}_r^0$ or $\hat{\bar{Y}}_r^{kw}$). This result is due to the large selection or nonresponse bias reduction by the proposed estimate $\hat{\bar{Y}}_r^{kwNR}$.

Table 2. Variance Ratio (VR) of the kwNR-weighted estimator and CV of NR-adjusted KW pseudoweights (kwNR) with varying $\mathrm{CV(KW)}$

| CV(KW) | $p(Y=1)=38\%$ | $p(Y=1)=87\%$ |
|---|---|---|
| | Variance Ratio (VR) | |
| Small (=0.52) | 1.03 | 1.04 |
| Large (=2.55) | 1.17 | 1.06 |
| | CV(kwNR) | |
| Small (=0.52) | 0.87 | 0.87 |
| Large (=2.55) | 3.78 | 3.74 |

Table 2 shows VR's of the kwNR estimates, i.e., the ratio of the mean of TL variance estimates divided by the empirical variance over 2,000 simulation runs, and CV of NR-adjusted KW weights for respondents with varying kernel weights and disease prevalences. It can be observed that the proposed TL estimates generally perform well with VR's close to one. When CV(KW) = 2.55 the TL variance tends to overestimate the variance with VR=1.17. This could be due to the small sample simulation error. As expected, CV(kwNR) tends to be larger than CV(KW) due to the extra variability that comes from estimating the response propensities.

In summary, via simulation studies, it is observed that the kwNR-weighted estimates from follow-up samples are approximately unbiased with the smallest MSE compared to the unweighted or the KW-weighted estimates. The proposed TL variance generally approximates the variance well with the variance ratios close to the value of one.

4. **Real Data Analysis** (to be added)

## 5. Conclusion and Discussion

In this paper, we present a statistical method to make population-based inferences using nonprobability longitudinal survey samples which are subject to both selection bias and nonresponse bias. We construct PS-based KW for the baseline nonprobability sample, then create nonresponse-adjusted KW pseudoweights (kwNR) for the follow-up nonprobability samples. The Taylor Linearization (TL) variance estimator of kwNR-weighted sample means is developed, which accounts for the variability due to estimating both the participation and response propensities. The proposed methods are evaluated via simulation studies and applied to the SARS-CoV-2 antibody seropositivity longitudinal study. The simulations studies showed kwNR-weighted estimates from follow-up samples are approximately unbiased and have the smallest MSE as compared to the unweighted or the KW-weighted estimates. The proposed TL variance generally approximates the variance well with the variance ratios close to the value of one. We apply our method to the SARS-CoV-2 antibody anti-Spike/RBD seropositivity longitudinal study.

There are limitations with this study. *First*, in the simulation study we generated the probability sample by simple random sampling (SRS). Although this was representative of the population, SRS doesn't mimic the complex sampling used to collect data from the large-scale national surveys such as the BRFSS probability sample. The simulation study with probability samples generated by complex sampling would also produce estimates $\hat{\bar{Y}}_r^{kwNR}$ that are approximately unbiased but with inflated variance from repeated complex sampling. *Second*, in the SARS-CoV-2 longitudinal study, we estimate anti-spike/RBD and anti-nucleocapsid seropositivity rates to illustrate the kwNR-weighted method ($\hat{\bar{Y}}_r^{kwNR}$), assuming the diagnostic test would have both high sensitivity (the proportion of people testing positive who have the disease) and high specificity (the proportion of people testing negative who do not have the disease). To report accurate estimates, sensitivity and specificity adjustments to $\hat{\bar{Y}}_r^{kwNR}$ should be conducted (Klumpp-Thomas et al., 2021). *Third*, In the method section, we adopt the participation propensity model from Wang et al. (2021) to estimate KW pseudoweights and the response propensity model from Little and Rubin (2002) to reduce nonresponse bias assuming data are missing at random.

Other nonresponse adjustment methods could be explored such as machine learning, weighting class methods of Little 1986 and calibration methods of Deville and Sarndal 1992. Readers interested in different nonresponse adjustment techniques are referred to the comprehensive review papers (Kalton and Flores-Cervantes, 2003; Brick, 2013). Lastly, in the real data application, we conducted a backward variable selection procedure to finalize the propensity and the response models. Further research is needed in choosing the best set of auxiliary variables to be included in either participation or response propensity models, especially when many variables are available.

**Disclosures:**

The views presented in this article are those of the authors and should not be viewed as official opinions or positions of the NIH, or U.S. Department of Health and Human Services. This work was supported in part by the intramural funds of the NIH, National Institute for Biomedical Imaging and Bioengineering, National Institute of Allergy and Infectious Diseases, National Center for the Advancement of Translational Sciences, and the National Cancer Institute.